\documentclass[aps,prd,draft,showpacs]{revtex4}

\begin{document}

\title{Renormalization Group Functions for the Radiative Symmetry Breaking Scheme}
\author{Chungku Kim}
\affiliation{Department of Physics, Keimyung University, Daegu 705-701, KOREA}
\date{\today}

\begin{abstract}
We obtain the renormalization group(RG) functions for the massless scalar
field theory where symmetry breaking occurs radiatively. After obtaining the
effective potential for the radiative symmetry breaking scheme from that of the 
minimal subtraction(MS) scheme by finite transformations for the classical field
and coupling constant, we calculate the corresponding change of the RG functions.
\end{abstract}
\pacs{11.15.Bt, 12.38.Bx}   
\maketitle

\affiliation{Department of Physics, College of Engineering, Keimyung
University, Daegu 705-701, KOREA}

\setcounter{page}{1}

In the radiative symmetry breaking scheme\cite{CW}, the scalar field does
not have a tree level mass and the spontaneous symmetry breaking occurs
radiatively from the effective potential\cite{EP}. The
electroweak(EW) model where the symmetry breaking occurs radiatively have
been studied extensively\cite{CWHiggs} due to its predictive power for the
magnitude of the Higgs boson mass. In order to investigate the higher-order
perturbative corrections in case of the large Higgs self-coupling, we need
the RG functions for the radiative symmetry breaking scheme. Recently, it was
conjectured\cite{conjec} that the RG functions for the radiative symmetry
breaking scheme (${\beta }_{R}$) can be obtained from that of
the minimal subtraction scheme ($\beta _{MS})$ as 
\begin{equation}
{\beta }_{R}(\lambda ) \stackrel{?}{=} \frac{\beta _{MS}(\lambda )}{1-\frac{\beta
_{MS}(\lambda )}{2\lambda }}
\end{equation}
and 
\begin{equation}
\gamma_{R}(\lambda )\stackrel{?}{=} \frac{\gamma _{MS}(\lambda )}{1-\frac{\beta
_{MS}(\lambda )}{2\lambda }}
\end{equation}

In this paper, we will investigate the relation between $\beta _{R}$ and $%
\beta _{MS}$ systematically. Starting from the effective potential for the
MS scheme whose renormalization functions are known, we
will obtain the effective potential for the radiative symmetry breaking
scheme by finite trnasformations of the field and coupling constant. Then we
will calculate corresponding change of the renormalization functions. For
simplicity, we will consider the case of the $\lambda  \phi ^{4}$ theory and
we will see that the above conjecture need corrections if the loop orders are
higher than three.

The classical Lagrangian of the massless $\lambda \phi ^{4}$ model is given
by 
\begin{equation}
L=\frac{1}{2}(\partial _{\mu }\phi _{B})^{2}+\frac{\lambda _{B}}{24}\phi
_{B}^{4}.
\end{equation}
where the bare field $\phi _{B}$ and the bare coupling constant $\lambda
_{B}.$ The effective potential of the massless $\lambda \phi ^{4}$ in the MS
scheme has the form 
\begin{equation}
V_{MS}(\lambda ,\phi ,\mu )=\sum_{l=0}^{\infty }\kappa ^{l}\lambda
^{l+1}\phi ^{4}\sum_{n=0}^{l}a_{l,n}L_{MS}^{n}\text{ }
\end{equation}
where $\kappa =(16\pi ^{2})^{-1}$, $a_{0,0}=\frac{1}{24}$ , $l$ is the
order of the loop and 
\begin{equation}
L_{MS}\equiv \log \left( \frac{\lambda \phi ^{2}}{2\mu ^{2}}\right) .
\end{equation}
Since the effective potential of the massless $\lambda \phi ^{4}$ model is
independent of the renormalization mass scale $\mu ,$ it satisfies the
renormalization group equation

\begin{equation}
\lbrack \mu \frac{\partial }{\partial \mu }+\beta _{MS}(\lambda )\frac{%
\partial }{\partial \lambda }+\gamma _{MS}(\lambda )\phi \frac{\partial }{%
\partial \phi }]V_{MS}(\lambda ,\phi ,\mu )=0
\end{equation}
where the coefficients of the RG functions $\beta _{MS}$ and $\gamma _{MS}$
are given by\cite{RGfn}

\begin{equation}
\beta _{MS}(\lambda )=\mu \frac{d\lambda }{d\mu }=\sum_{l=1}^{\infty }\kappa
^{l}\lambda ^{l+1}\text{ }b_{l}=3\kappa \lambda ^{2}-\frac{17}{3}\kappa
^{2}\lambda ^{3}+(\frac{145}{8}+12\text{ }\varsigma (3))\kappa ^{3}\lambda
^{4}+\cdot \cdot \cdot
\end{equation}
and

\begin{equation}
\gamma _{MS}(\lambda )=\frac{\mu }{\phi }\frac{d\phi }{d\mu }%
=\sum_{l=2}^{\infty }\kappa ^{l}\lambda ^{l+1}\text{ }g_{l}=-\frac{1}{12}%
\kappa ^{2}\lambda ^{2}+\frac{1}{16}\kappa ^{3}\lambda ^{3}+\cdot \cdot \cdot.
\end{equation} The effective potential for the $\lambda 
\phi ^{4}$ theory in the MS scheme was obtained up to two-loop order\cite
{2loop}. In case of the three-loop order, we can determine terms which
depend on logarithms by using the RG improvement of the
effective potential\cite{run}. As a result, we obtain 
\begin{eqnarray}
V_{MS}(\lambda ,\phi ,\mu ) &=&\phi ^{4}[\frac{1}{24}\lambda +\kappa \lambda
^{2}(\frac{1}{16}L_{MS}-\frac{3}{32})+\kappa ^{2}\lambda ^{3}(\frac{3}{32}%
L_{MS}^{2}-\frac{5}{16}L_{MS}+\frac{11}{32}+\frac{1}{2}\Omega (1))  \nonumber
\\
&&+\kappa ^{3}\lambda ^{4}(\frac{9}{64}L_{MS}^{3}-\frac{143}{192}L_{MS}^{2}+(%
\frac{701}{384}+\frac{\text{ }\varsigma (3)}{4}+\text{ }\frac{9}{4}\Omega
(1))L_{MS}+a_{30})+\text{ }\cdot \cdot \cdot ]
\end{eqnarray}
where 
\begin{equation}
\Omega (1)=-\frac{1}{2\sqrt{3}}\sum_{n=1}^{\infty }\frac{1}{n^{2}}\sin (%
\frac{n\pi }{3})\simeq -0.293\text{.}
\end{equation} and $a_{30}$ is a constant.
It is easy to see that $V_{MS}(\lambda ,\phi ,\mu )$ given in (9) satisfies
the RG equation (6) up to the order $O(\kappa ^{3})$ and $a_{30}$ do not
contribute up to this order.

\subsection{Two Step Procedure}

In this procedure, we will follow two steps to transform the
effective potential for the minimal subtraction $V_{MS} (\lambda ,\phi ,\mu )
$ into the effective potential for the radiative symmetry breaking and will
investigate the corresponding change of the RG functions.

STEP I: In this step, by making a transformation $\mu ^{2}\rightarrow $ $\mu
^{2}\lambda /2 $, we change the logarithms $L_{MS}$ in $V_{MS} ( \lambda
,\phi ,\mu )$ to $\widetilde{L}$ where $\widetilde{L}\equiv \log \left( \frac{\phi ^{2}}{\mu ^{2}}%
\right)$. As a result, we obtain $\widetilde{V}(\lambda ,\phi ,\mu )$ where 
\begin{eqnarray}
\widetilde{V}(\lambda ,\phi ,\mu ) &=& \sum_{l=0}^{\infty }\kappa
^{l}\lambda ^{l+1}\phi ^{4}\sum_{n=0}^{l}a_{l,n}\widetilde{L}^{n}  \nonumber \\
&=& \phi ^{4}[\frac{1}{24}\lambda +\kappa \lambda ^{2}(\frac{1}{16}\widetilde{L}-\frac{3%
}{32})+\kappa ^{2}\lambda ^{3}(\frac{3}{32}\widetilde{L}^{2}-\frac{5}{16}\widetilde{L}+\frac{11}{32}+%
\frac{1}{2}\Omega (1))  \nonumber \\
&&+\kappa ^{3}\lambda ^{4}(\frac{9}{64}\widetilde{L}^{3}-\frac{143}{192}\widetilde{L}^{2}+(\frac{701%
}{384}+\frac{\text{ }\varsigma (3)}{4}+\text{ }\frac{9}{4}\Omega
(1))\widetilde{L}+a_{30})+\cdot \cdot \cdot ]
\end{eqnarray}
Let us emphasize the fact that both $V_{MS}(\lambda ,\phi ,\mu )$ and $%
\widetilde{V}(\lambda ,\phi ,\mu )$ contains the same coefficient $a_{l,n}$.
It is known that\cite{Jones1} the $\widetilde{V}(\lambda ,\phi ,\mu )$
satisfies the RG equation
\begin{equation}
\lbrack \mu \frac{\partial }{\partial \mu }+\widetilde{\beta }\frac{\partial 
}{\partial \lambda }+\widetilde{\gamma }\phi \frac{\partial }{\partial \phi }%
]\widetilde{V}(\lambda ,\phi ,\mu )=0
\end{equation}
where 
\begin{equation}
\widetilde{\beta }(\lambda )=\frac{\beta _{MS}(\lambda )}{1-\frac{\beta
_{MS}(\lambda )}{2\lambda }}=\sum_{l=1}^{\infty }\kappa ^{l}\lambda ^{l+1}%
\text{ }\widetilde{g}_{l}
\end{equation}
and 
\begin{equation}
\widetilde{\gamma }(\lambda )=\frac{\gamma _{MS}(\lambda )}{1-\frac{\beta
_{MS}(\lambda )}{2\lambda }}=\sum_{l=1}^{\infty }\kappa ^{l}\lambda ^{l+1}%
\text{ }\widetilde{b}_{l}
\end{equation}
By expanding (13) and (14), we obtain $\widetilde{b}_{l}$ and $\widetilde{g}%
_{l}$: 
\begin{eqnarray}
\text{ }\widetilde{b}_{1} &=&b_{1}=3 \\
\text{ }\widetilde{b}_{2} &=&b_{2}+\frac{1}{2}b_{1}^{2}=-\frac{7}{6} \\
\text{ }\widetilde{b}_{3} &=&b_{3}+b_{1}b_{2}+\frac{1}{4}b_{1}^{3}=\frac{63}{%
8}+12\text{ }\varsigma (3) \\
\text{ }\widetilde{g}_{2} &=&g_{2}=-\frac{1}{12} \\
\text{ }\widetilde{g}_{3} &=&g_{3}+\frac{1}{2}g_{2}b_{1}=-\frac{1}{16}
\end{eqnarray}
Then one can easily see that $\widetilde{V}(\lambda ,\phi ,\mu )$ given in
(11) satisfies the RG equation (12) up to the order $O(\kappa ^{3})$. In
order to see that the the RG equation (12) is satisfied to all orders, let us
first substitute $V_{MS}(\lambda ,\phi ,\mu )$ given in (4) to the RG
equation (6). As a result, we obtain 
\begin{equation}
\sum_{l=0}^{\infty }\kappa ^{l}\lambda ^{l+1}\phi
^{4}\sum_{n=1}^{l}[-2na_{l,n}+\frac{\beta _{MS}(\lambda )}{\lambda }%
\{(l+1)a_{l,n-1}+na_{l,n}\}+\gamma
_{MS}\{4a_{l,n-1}+2na_{l,n}\}]L_{MS}^{n-1}=0
\end{equation}
If we substitute the perturbative expansions given in Eqs.(7) and (8) into
the above equation, we can determine the coefficients $a_{l,n}$. Since the
$L_{MS}$ in (20) is independent of $a_{l,n}$, we can replace it with any quantity
which is independent of $a_{l,n}$. Then, by replacing $L_{MS}$
with $\widetilde{L}$ in (20) and by arranging terms, we obtain 
\begin{equation}
\sum_{l=0}^{\infty }\kappa ^{l}\lambda ^{l+1}\phi ^{4}\sum_{n=1}^{l}[-(1-%
\frac{\beta _{MS}(\lambda )}{2\lambda })2na_{l,n}+\frac{\beta _{MS}(\lambda )%
}{\lambda }(l+1)a_{l,n-1}+\gamma _{MS}\{4a_{l,n-1}+2na_{l,n}\}]\widetilde{L}^{n-1}=0
\end{equation}
By using (11), we can write above equation as 
\begin{equation}
\lbrack (1-\frac{\beta _{MS}(\lambda )}{2\lambda })\mu \frac{\partial }{%
\partial \mu }+\beta _{MS}(\lambda )\frac{\partial }{\partial \lambda }%
+\gamma _{MS}(\lambda )\phi \frac{\partial }{\partial \psi }]\widetilde{V}%
(\lambda ,\phi ,\mu )=0.
\end{equation}
and by dividing this equation with $(1-\frac{\beta _{MS}(\lambda )}{2\lambda 
})$, we obtain (12).

STEP II: In order to transform the $\widetilde{V}
(\lambda ,\phi ,\mu )$ into the effective potential for the radiative symmetry
breaking scheme $V_R$, we need to add finite counterterms to satisfy given
renormalization condition. As in case of the infinite counterterms, these
finite counterterms will also contribute to the higher order loop expansion
and this amounts to the redefinition of the coupling constants\cite{Jones2}.
Then, by defining new coupling constant $\eta$ and the classical field $\psi$ as
\begin{equation}
\lambda (\eta )=\sum_{l=0}^{\infty }\kappa ^{l}c_{l}(\eta )\text{ }(%
\text{ }c_{0}=1)
\end{equation}
and 
\begin{equation}
\phi (\psi ,\eta )=\psi \sum_{l=0}^{\infty }\kappa ^{l}d_{l}(\eta )
\text{ }(\text{ }d_{0}=1)
\end{equation}
we can obtain the effective potential for the radiative symmetry breaking $%
V_{R}(\eta ,\psi ,\mu )$ as
\begin{equation}
V_{R}(\eta ,\psi ,\mu )=\widetilde{V}(\lambda (\eta ),\phi (\psi ,\eta ),\mu
)
\end{equation}
which should satisfy the given renormalization condition 
\begin{equation}
\left[ \frac{d^{4}V_{R}(\eta ,\psi ,\mu )}{d\psi ^{4}}\right] _{\psi =\mu
}=\eta.
\end{equation}

The constants $c_{l}$ and $d_{l}$ can be determined order by order from the
renormalization condition given in (25). Actually, we obtain
\begin{eqnarray}
c_{1} &=&-4\eta^2 , \text{ } \text{ } d_{1}=0 \\
c_{2} &=&(\frac{115}{4}-12\Omega (1))\eta^3, \text{ } \text{ } d_{2}=0 \\
c_{3} &=&(-\frac{28465}{96}+15\Omega (1)-25\varsigma (3)-24a_{30})\eta^4, 
\text{ } \text{ } d_{3}=0.
\end{eqnarray}
Althouh $d_{i}=0$ in this step, it turns out that $d_{i} \neq 0$ in case of the
one step procedure (see(40) and (41)).
Then $\beta _{R}(\eta )$ can be obtained as 
\begin{equation}
\beta _{R}(\eta )\equiv \mu \frac{\partial \eta }{\partial \mu }=\frac{%
\partial \eta }{\partial \lambda }\mu \frac{\partial \lambda }{\partial \mu }%
=\frac{\widetilde{\beta }(\lambda (\eta ))}{\frac{\partial \lambda (\eta )}{%
\partial \eta }}.
\end{equation}
In order to obtain $\gamma _{R}(\eta )$, let us invert the
Eq.(24) as 
\begin{equation}
\psi (\phi ,\eta )=\phi [1-\kappa d_{1} +\kappa^2 (d_{1}^{2}-d_{2})
+\kappa^3 (-d_{3}+2d_{2}d_{1}-d_{1}^{3}) \cdot \cdot \cdot ]\equiv \phi
\sum_{l=0}^{\infty }\kappa ^{l}e_{l}(\eta).
\end{equation}
Then $\gamma _{R}(\eta )$ can be obtained as 
\begin{eqnarray}
\gamma _{R}(\eta ) &\equiv &\frac{\mu }{\psi }\frac{\partial \psi }{\partial
\mu }=\frac{1}{\psi }[\mu \frac{\partial \phi }{\partial \mu }%
\sum_{l=0}^{\infty }\kappa ^{l}e_{l}(\eta)+\phi \mu \frac{\partial \eta }{%
\partial \mu }\sum_{l=0}^{\infty }\text{ }\kappa ^{l}
\frac{\partial e_{l}(\eta)}{\partial \eta }] \nonumber \\
&=&\widetilde{\gamma}(\lambda (\eta ))+\beta _{R}(\eta )\sum_{l=0}^{\infty }\kappa
^{l}d_{l}(\eta)\sum_{m=0}^{\infty }\text{ }\kappa ^{m}\frac{\partial e_{l}(\eta)}
{\partial \eta }.
\end{eqnarray}
Note that the coefficients $c_{l}$ and $d_{l}$ determine the order $\kappa
^{l+1}$ terms of the RG functions. By substituting Eqs.(23),(24),(27),(28) and
(29) to Eqs.(23),(24) and(25), we obtain the effective potential and the RG
functions as
\begin{eqnarray}
V_{R}(\eta ,\psi ,\mu ) &=&\phi ^{4}[\frac{1}{24}\eta +\kappa \eta ^{2}(%
\frac{1}{16}L-\frac{25}{96})+\kappa ^{2}\eta ^{3}(\frac{3}{32}L^{2}-\frac{13%
}{16}L+\frac{55}{24})  \nonumber \\
&&+\kappa ^{3}\eta ^{4}(\frac{9}{64}L^{3}-\frac{359}{192}L^{2}+(\frac{3905}{%
384}+\frac{\text{ }\varsigma (3)}{4}+\text{ }\frac{3}{4}\Omega (1))L-(\frac{%
53845}{2304}+\frac{25}{8}\Omega (1)+\frac{25}{24}\varsigma (3)))+\text{ }%
\cdots ],
\end{eqnarray}
where $L\equiv \log \left( \frac{ \psi ^{2}}{\mu ^{2}}\right)$
\begin{eqnarray}
\beta _{R}(\eta ) &=&\kappa \widetilde{b}_{1}\eta ^{2}+\kappa ^{2}\widetilde{%
b}_{2}\eta ^{3}+\kappa ^{3}(\widetilde{b}_{3}+(c_{1}^{2}-c_{2})\widetilde{b}%
_{1}+c_{1}\widetilde{b}_{2})\eta ^{4}+\cdot \cdot \cdot \cdot \cdot  
\nonumber \\
&=&3\kappa \eta ^{2}-\frac{7}{6}\kappa ^{2}\eta ^{3}+(-\frac{617}{24}+12%
\text{ }\varsigma (3)+36\Omega (1))\kappa ^{3}\eta ^{4}+\cdot \cdot \cdot 
\end{eqnarray}
and 
\begin{eqnarray}
\gamma _{R}(\eta ) &=&\widetilde{g}_{2}\kappa ^{2}\eta ^{2}+(\widetilde{g}%
_{3}+2c_{1}\widetilde{g}_{2})\kappa ^{3}\eta ^{3}+\cdot \cdot \cdot  
\nonumber \\
&=&-\frac{1}{12}\kappa ^{2}\eta ^{2}+\frac{29}{48}\kappa ^{3}\eta ^{3}+\cdot
\cdot \cdot 
\end{eqnarray}
Note that $V_{R}(\eta ,\psi ,\mu )$ agrees with the N=1 case of the two loop
effective potential for the O(N) symmetric scalar field theory\cite{Jackiw}
and satisfies the RG equation

\begin{equation}
\lbrack \mu \frac{\partial }{\partial \mu }+\beta _{R}(\eta )\frac{\partial 
}{\partial \eta }+\gamma _{R}(\eta )\psi \frac{\partial }{\partial \psi }%
]V_{R}(\eta ,\psi ,\mu )=0
\end{equation}
up to the order $\kappa ^{3}$.

\subsection{One Step Procedure}

We can obtain the effective potential for the radiative symmetry breaking scheme
directly from the effective potential for the minimal subtraction $V_{MS}(\lambda ,\phi
,\mu )$ without using STEP.I of the two step procedure by using the finite
transformation of the coupling constant $\lambda$ and the classical field 
$\phi $ as given in Eqs.(23) and (24).
Then the effective potential for the radiative symmetry
breaking scheme $V_{R}(\eta ,\psi ,\mu )$ defined by
\begin{equation}
V_{R}(\eta ,\psi ,\mu )=V_{MS}(\lambda (\eta ),\phi (\psi ,\eta ),\mu )
\end{equation} 
should satisfy following conditions: 

(i) $V_{R}(\eta ,\psi ,\mu )$ should satisfy both the renormalization condition given in (26).

(ii)All the $\log ^{m}\left( \frac{\lambda
\phi ^{2}}{2\mu ^{2}}\right) (1\leq m\leq l)$ terms in the order $\kappa
^{l}$ terms of $V_{MS}$ should be transformed to $\log ^{m}\left( \frac{\psi ^{2}}{%
\mu ^{2}}\right)$.

As a result of the condition (ii), the order $\kappa ^{l}$ term of the coefficients
$c_{l}$ and $ d_{l}$ contains the terms depending on powers of  $\log \left( \frac{\eta }{2%
}\right) $. In the present case, the RG functions are given by
\begin{equation}
\beta _{R}(\eta )=\frac{\beta_{MS} (\lambda (\eta ))}{\frac{\partial \lambda (\eta )}
{\partial \eta }}.
\end{equation} and
\begin{eqnarray}
\gamma _{R}(\eta )=\gamma_{MS}(\lambda (\eta ))+\beta _{R}(\eta )\sum_{l=0}^{\infty }\kappa
^{l}d_{l}(\eta)\sum_{m=0}^{\infty }\text{ }\kappa ^{m}\frac{\partial e_{l}(\eta)}
{\partial \eta }.
\end{eqnarray}
instead of (30) and (32).
If we substitute the resulting coefficients to Eqs.(38) and (39), we can obtain
the RG functions $\beta _{R}(\eta )$ and $\gamma _{R}(\eta )$. It turns out that these two conditions
does not fix the coefficients $c_{l}$ and $d_{l}$ uniquely and some choices can lead to RG functions
depending on the $\log \left( \frac{\eta }{2}\right) $ terms. In order to fix the
coefficients $c_{l}$ and $d_{l}$ uniquely, we demand additional condition:

(iii) The resulting order $\kappa ^{l+1}$ terms of the RG
functions does not contains the $\log \left( \frac{\eta }{2}\right) $. 

Then, the coefficients $c_{l}$ and $ d_{l}$ can be obtained as 
\begin{eqnarray}
c_{1} &=& (-\frac{3}{2}\log \left( \frac{\eta }{2}\right) -4)\eta^2,   \\
c_{2} &=&(\frac{9}{4}\log ^{2}\left( \frac{\eta }{2}\right) +\frac{89}{6}\log
\left( \frac{\eta }{2}\right) +\frac{139}{4}-12\Omega (1))\eta^3,   \\
d_{1} &=&0   \\
d_{2} &=&\frac{1}{24} \eta^2 \log \left( \frac{\eta }{2}\right) .
\end{eqnarray}
By sustituting the coefficients $c_{l}$ and $d_{l}$ into Eqs. (37),(38) and
(39) we obtain the exactly same effective potential and RG functions as
in Eqs.(33),(34) and (35).

In summary, we have obtained the RG functions for the radiative symmetry breaking
scheme from those for the minimal subtraction scheme by two
different ways. One was by changing the mass scale $\mu $ followed by a
finite transformations of the coupling constants and classical field to
satisfy the renormalization condition. The other was by using only
the finite transformations of the coupling constants and classical field that depends
on logarithms of the coupling constants. We have
confirmed that he two different methods give
the same results for the effective potential and the RG functions and 
satisfy the RG equation up to three-loop order. The latter method can be used in
case of the theories which have two or more different mass scales say in $%
O(N)$ symmetric scalar field theory and scalar electrodynamics and this is in progress.

\center{ACKNOWLEDGMENTS} \endcenter

This research was supported by the Institute of the Basic Science Research
Center.

\end{document}